\documentclass[sigconf]{acmart}
\usepackage{soul}

\usepackage{tabularx}
\usepackage{array}
\PassOptionsToPackage{table}{xcolor}

\usepackage{amsmath}
\usepackage[utf8]{inputenc}
\usepackage{enumitem}
\usepackage{hyperref}
\usepackage{url}

\usepackage{geometry}
\usepackage{subcaption}
\usepackage{eurosym}

\AtBeginDocument{%
  }

\copyrightyear{2025}
\acmYear{2025}
\setcopyright{rightsretained}
\setcctype{CC-BY}
\acmConference[TEI '25]{Nineteenth International Conference on Tangible, Embedded, and Embodied Interaction}{March 4--7, 2025}{Bordeaux / Talence, France}
\acmBooktitle{Nineteenth International Conference on Tangible, Embedded, and Embodied Interaction (TEI '25), March 4--7, 2025, Bordeaux / Talence, France}\acmDOI{10.1145/3689050.3704945}
\acmISBN{979-8-4007-1197-8/25/03}

\begin{document}

\title[Narrowing the Digital Divide for Older Adults]{``It's Like Not Being Able to Read and Write'':\\Narrowing the Digital Divide for Older Adults and Leveraging the Role of Digital Educators in Ireland}

\author{Melanie Gruben}
\orcid{0009-0008-9332-5205}
\affiliation{%
  \institution{Munster Technological University}
  \city{Cork}
  \country{Ireland}
}
\email{melanie.gruben@mtu.ie}

\author{Ashley Sheil}
\orcid{0000-0001-7750-9495}
\affiliation{%
  \institution{Munster Technological University}
  \city{Cork}
  \country{Ireland}
  }  
\email{ashley.sheil@mtu.ie}

\author{Sanchari Das}
\orcid{0000-0003-1299-7867}
\affiliation{%
  \institution{University of Denver}
  \city{Denver, Colorado}
  \country{USA}}
\email{sanchari.das@du.edu}

\author{Michelle O Keeffe}
\orcid{0009-0008-7130-9941}
\affiliation{%
  \institution{Munster Technological University}
  \city{Cork}
  \country{Ireland}
}
\email{michelle.okeeffe2@mtu.ie}

\author{Jacob Camilleri}
\orcid{0009-0009-4912-258X}
\affiliation{%
  \institution{Munster Technological University}
  \city{Cork}
  \country{Ireland}
}
\email{jacob.camilleri@mtu.ie}

\author{Moya Cronin}
\orcid{0009-0008-2831-186X}
\affiliation{%
  \institution{Munster Technological University}
  \city{Cork}
  \country{Ireland}
}
\email{moya.cronin@mycit.ie}

\author{Hazel Murray}
\orcid{0000-0002-5349-4011}
\affiliation{%
 \institution{Munster Technological University}
 \city{Cork}
 \country{Ireland}}
 \email{hazel.murray@mtu.ie}

\renewcommand{\shortauthors}{   Gruben et al.}

\begin{abstract}
As digital services increasingly replace traditional analogue systems, ensuring that older adults are not left behind is critical to fostering inclusive access. This study explores how digital educators support older adults in developing essential digital skills, drawing insights from interviews with $34$ educators in Ireland. These educators, both professional and volunteer, offer instruction through a range of formats, including workshops, remote calls, and in-person sessions. Our findings highlight the importance of personalized, step-by-step guidance tailored to older adults' learning needs, as well as fostering confidence through hands-on engagement with technology. Key challenges identified include limited transportation options, poor internet connectivity, outdated devices, and a lack of familial support for learning. To address these barriers, we propose enhanced public funding, expanded access to resources, and sustainable strategies such as providing relevant and practical course materials. Additionally, innovative tools like simulated online platforms for practicing digital transactions can help reduce anxiety and enhance digital literacy among older adults. This study underscores the vital role that digital educators play in bridging the digital divide, creating a more inclusive, human-centered approach to digital learning for older adults.
\end{abstract}



\maketitle

\section{Introduction}
By 2030, the number of adults over $65$ will exceed that of children aged 0-9 for the first time, according to the United Nations~\cite{UN-report}. This significant demographic shift highlights the urgent need to establish comprehensive systems that support older adults in adapting to a rapidly digitising world. The World Health Organisation (WHO) defines active ageing as the process of optimising opportunities for health, security, and participation to enhance the quality of life as people age~\cite{WHO-active-ageing}. Central to this concept are autonomy and independence: autonomy refers to the perceived ability to control, cope with, and make personal decisions about daily life according to one's preferences, while independence is the ability to perform daily living functions and live in the community with minimal assistance~\cite{WHO-active-ageing}. Digital skills are crucial in helping older adults maintain engagement, connection, and independence, thereby enriching their lives with new opportunities~\cite{millward2003grey}.

For the purpose of the present study, we operationalise ``digital skills'' in accordance with the United Kingdom government's Essential Digital Skills Framework~\cite{uk2019essential}. This framework defines digital skills as the ability to be ``safe, legal, and confident online,'' which encompasses a range of competencies. These competencies include problem-solving, communicating, transacting, and handling information and content. We chose this framework due to its comprehensive nature, as it captures not only the technical abilities required to navigate digital environments but also the essential safety and legal considerations, which were highly relevant to our study participants' concerns, particularly with regard to cybersecurity. This alignment allowed us to examine a broad spectrum of skills, from basic navigation and communication tasks to more complex issues such as managing digital transactions and safeguarding personal information online. By adopting this operational definition, we ensure that the scope of ``digital skills'' is clearly delineated and applicable to the experiences reported by both educators and learners in the context of our research. In this way, our use of the framework provides a structured lens for analyzing the varying levels of digital competence among older adults, and highlights both the foundational skills required for digital inclusion and the barriers that arise from gaps in these skills.

Given that digital engagement increasingly shapes our everyday interactions with essential services, from healthcare to banking, there is an emerging need to view digital literacy as a fundamental life skill. In this context, the process of learning digital skills can be seen as an embedded and embodied experience, where older adults must navigate complex technological landscapes and integrate these tools into their daily routines. The Digital Economy and Society Index (DESI), a European measure of digital competences, reported that in 2021, 47\% of adults in Ireland lacked basic digital skills~\cite{DESI}. This digital divide risks exacerbating social isolation and exclusion among older adults. The Centre for Ageing Better has noted that ``digital exclusion will become less about whether you are online and more about what you are doing online''~\cite{centreforagingbetter}, emphasising the importance of digital literacy in today's society.

Digital educators play a vital role in bridging this digital divide. These educators provide technical support and digital skills instruction to older adults, working on a volunteer, contract, or permanent basis. They deliver a variety of services, including remote on-call support, in-person drop-in sessions, workshops, and classroom-based instruction. Digital educators are employed by a range of organisations such as charities, libraries, government bodies, colleges, and community centres. Unlike family members who might provide sporadic tech support, digital educators offer structured, context-rich, and step-by-step guidance that empowers older adults and enhances their digital competency. Through tactile and face-to-face interaction, educators help older adults engage with technology on a deeper level, fostering hands-on learning experiences that are crucial for developing long-term digital fluency. In Ireland, the role of digital educators is particularly critical due to the country's significant immigrant population~\cite{gilmartin2018immigrant}. Immigration often splits families, leaving older adults without immediate familial support for technological issues. This context makes the support provided by digital educators indispensable for many older adults in Ireland.

However, the provision of digital education services faces numerous challenges. Public resources are dwindling, global crises are growing, and societal values are shifting, leading to the deprioritisation of older adults~\cite{davies2023healthcare, payne2022ageism, minichiello2000perceptions}. More robust and expanded services are needed across all areas of care for older adults. With everyday services such as banking, healthcare, and government interactions transitioning to digital platforms, the lack of adequate digital skills can leave older adults vulnerable to exclusion and isolation. Without these skills, they may struggle to navigate essential services or recognize online threats, which could have serious implications for their safety and well-being. Recognising the critical importance of these services, we conducted a study to identify the elements that make digital skills programmes most effective for older adults. We aimed to understand the systemic factors that influence older adults' ability to learn digital skills and to explore how the support framework for digital educators can be strengthened and advanced. 

Our study takes a close look at the embodied experiences of digital educators in Ireland, focusing on how they support older adults through both tangible and intangible means. Through semi-structured interviews with $34$ educators, we explore how they not only provide technical instruction but also help build confidence and reduce the anxiety older adults often feel towards technology. We took special interest in Ireland due to its economically deprived history~\cite{bradley1999history}, which means that many Irish older people may still be adjusting to using technology. We posit, then, that Irish digital educators may have an especially challenging task and may provide unique insights on the topic. We hypothesise that Irish digital educators face particularly challenging tasks and can provide valuable insights into the effective delivery of digital skills education. Our research focused on answering the following research questions through semi-structured interviews:

\begin{itemize}
    \item[\textbf{RQ1}] What specific factors in the design and implementation of digital skills programmes maximise learning outcomes for older adults?
    \item[\textbf{RQ2}] What individual-level barriers do older adults face in learning digital skills and cyber security?
    \item[\textbf{RQ3}] How do systemic issues, including policy, access to technology, and socioeconomic factors contribute to the addressing or worsening of the digital divide among older adults?
    \item[\textbf{RQ4}] How can digital educators be better equipped and supported through training, resources, and institutional backing to effectively teach digital skills to older adults?
\end{itemize}

\textit{Contributions:} This study provides a comprehensive exploration of the role and impact of digital educators in fostering digital skills development among older adults. The challenges identified in the interviews include fear and anxiety around technology, inaccessible technical language, limited transportation to classes, poor internet connectivity, and financial barriers related to updating devices. On the other hand, opportunities for innovative educational tools and methods have emerged. For instance, we recommend that institutions create simulated online banking platforms, enabling older adults to practice digital banking in a safe and controlled environment, helping reduce fear and build confidence. Our research provides empirical data to guide policymakers on the critical need for state funding and incentives to support digital skills education for older adults. By identifying the systemic factors affecting digital learning, including policy and socioeconomic issues, we advocate for a more inclusive and supportive framework, laying the groundwork for developing targeted interventions. Ultimately, this study contributes to the broader understanding of how interaction design and tangible teaching approaches can enhance the digital learning experiences of older adults. It offers actionable recommendations to support sustainable development in digital literacy, ensuring that older adults remain actively engaged and independent in a digital-first society.

\section{Background \& Related Work}
\subsection{Importance of Older Adults Gaining Digital Skills}
The COVID-19 pandemic heightened the necessity of online participation in society, which in turn exacerbated the digital divide in terms of access and digital skills~\cite{wilson2023learning,van2022zooming,harley2022together}. Pihlainen et al. examined the motivations behind why older adults engage in digital skills training in Austria, Finland, and Germany, highlighting practical necessity, managing finances, healthcare, and staying connected with family and friends as key drivers~\cite{pihlainen2023older}. Our study expands on these findings by investigating these motivations among older adults in Ireland. Health is a critical motivator for older adults to learn digital skills. Cho et al. discovered that regular internet users experienced approximately half the risk of dementia compared to non-users, with long-term internet use in late adulthood associated with delayed cognitive impairment~\cite{cho2023internet}. 

In Ireland, accessing healthcare presents additional challenges. According to a survey by the Irish Independent, more than two-thirds (66\%) of general practitioners in rural Ireland are unable to accept new patients, with some reporting appointment wait times of up to two weeks~\cite{Irish-Indo-gps}. The 2024 roadmap for eHealth Ireland, launched by the Health Service Executive (HSE), aims to transform healthcare delivery through digital technology and information systems~\cite{hse-telehealth}. This initiative includes remote consulting, necessitating that Irish older adults be capable and comfortable with using digital tools for healthcare interactions, to which this study hopes to contribute. 

The COVID-19 pandemic underscored the advantages of online shopping and socialising~\cite{mouratidis2021covid,chen2022workshops,zhang2024tablecanvas,weijdom2022performative}, but it also revealed that older adults struggled due to a lack of digital skills~\cite{digital-age-action,fuchsberger2024remote}. Research indicates a 38\% probability of experiencing a pandemic like COVID-19 in one's lifetime~\cite{marani2021intensity}, suggesting that future lockdowns are possible. Older adults, being at higher risk from viruses~\cite{chen2021aging}, will benefit from the ability to conduct their shopping online. During the COVID-19 lockdown in Ireland, the Irish ALONE national support line received $26,174$ calls between March 9th and July 5th, 2020. Of these calls, 55\% were from individuals over 70, and 75\% of the callers were living alone~\cite{trinity-press}, highlighting the importance of digital connectivity for maintaining social ties when in-person interactions are not feasible due to infection fears or mobility issues.

With the closure of many physical banks in Ireland, older adults feel pressured to transition to online banking~\cite{finance-older}. This poses significant challenges, including fear of financial scams, lack of digital skills, and insufficient support~\cite{finance-older, thomas2023exploring}. To fully benefit from the expanding digital age, older adults must bridge the digital gap or risk being left behind~\cite{mubarak2022elderly}. To achieve this, policymakers and stakeholders must prioritise digital literacy initiatives tailored to older adults to foster inclusivity and ensure that they are not excluded from the advantages of digital advancements.

\subsection{Cyber Scams \& Their Impact on Older Adults}
Although older adults are not necessarily targeted more frequently than younger individuals, scams targeting older adults is increasing \cite{yu2023vulnerability, aarp-scam} and factors such as cognitive decline and inherent trust contribute to their vulnerability~\cite{shang2022psychology}. Therefore, fraud prevention strategies must be carefully designed to address these unique vulnerabilities. In 2022, there were 88,262 fraud complaints from older adults, resulting in \$3.1 billion in losses. These scams often involve government impersonation, sweepstakes, and robocalls, leaving victims with little to no recourse to recover their losses~\cite{NCO}. Approximately 5.4\% of cognitively intact older adults in the U.S. fall victim to financial fraud annually, highlighting the urgent need for targeted prevention strategies~\cite{burnes2017prevalence}. 
Furthermore surveys conducted from 2021 to 2023 indicate that 75\% of adults aged 50-80 encountered a scam attempt, with 30\% experiencing actual fraud. Those in poorer health are particularly at risk, and many older adults express a need for more information on how to protect themselves from scams~\cite{MichiganUni}. Older adults often rely on family and friends for cybersecurity advice rather than online resources and show a preference for broadcast media over internet sources for such information~\cite{nicholson2019if}. Their rapid adoption of digital technologies during the COVID-19 pandemic further underscores the necessity for tailored cybersecurity solutions~\cite{morrison2023recognising}.

Phishing has also become more pervasive, impacting various demographics with significant financial and emotional consequences \cite{das2019all}. With an estimated $3.4$ billion spam emails sent daily, Google Mail alone blocks around $100$ million phishing emails daily. In 2022, over 48\% of all emails were classified as spam, with Millennials and Gen-Z users notably susceptible to phishing attacks~\cite{AAG}. Privacy and data security concerns also play a significant role in older adults' acceptance of online services. While they employ various protection strategies, there is a strong demand for greater support from authorities and online service providers~\cite{ellefsen2022privacy}.

Many older adults experience frustration or resignation towards privacy attacks, which can lead to avoidance of technology altogether~\cite{ray2019woe}. As noted at the beginning of this section, contrary to perceptions, younger population are more frequently targeted by scams; however, older individuals tend to be more diligent in monitoring their bank accounts and taking preventive measures~\cite{ptsb, FraudSmart}. 
Despite their due diligence, financial fraud has a more severe impact on older adults, with higher average amounts stolen compared to younger individuals. The emotional and physical impacts of fraud on older adults necessitate support programmes that address not only financial but also the non-financial impacts of such crimes~\cite{kemp2023consumer}. Thus, understanding the cyber struggles of the older adult is critical yet understudied especially for the population in Ireland.

\subsection{Barriers to Older Adults Attaining Digital Skills}
Numerous studies involving older adults highlight lack of confidence and fear as significant barriers to attaining digital skills ~\cite{beh2018achieving, di2019psychological, holgersson2021cybersecurity, wilson2023understanding, feng2023understanding}. Another psychological barrier comes in the form of internalised ageism \cite{kottl2021but, barrie2021because}. These in turn can translate into a reluctance to learn and lack of affinity to technology~\cite{arthanat2019multi}. 
Additional obstacles include physical and cognitive limitations, limited access to education, outdated technology, and financial constraints. These obstacles collectively contribute to what has been termed the `grey divide'~\cite{millward2003grey}. 
Physical and cognitive limitations are common as people age and as such, 
age-related changes such as vision, hearing loss and fine motor difficulties can pose barriers to learning digital skills~\cite{gitlow2014technology, bhattacharjee2020older, butt2023barriers}. Other barriers include difficulties in understanding manuals, technical terms and limited reading skills, especially for those with intellectual disabilities~\cite{schlomann2022older}.

To stay current with digital skills, having up-to-date equipment is essential. This is not always possible for older adults~\cite{mubarak2022elderly, tomczyk2023barriers}. 
Harris et al. highlight that the cost of smart technology is a substantial financial barrier to learning digital skills. They also note difficulties in understanding and using technology, as well as privacy concerns as additional obstacles~\cite{harris2022older}. 
Studies similarly point out financial constraints, lack of family support, and limited access to computers and the internet due to low-income levels, as critical impediments~\cite{mckee2006older, friemel2016digital}. Due to the fact that Ireland has a considerable urban-rural divide and a poor public transport system ~\cite{transport-justice, EU-Barometer}, attending in person digital classes can pose an additional barrier for Irish older adults \cite{flynn2024keeping}.  Another significant barrier for older Irish adults, is internet connectivity in rural areas, with Ireland lagging behind on the promised widespread broadband rollout~\cite{Irish-Indo-Broadband, EU-Barometer}.
Overcoming these barriers necessitates a combination of customised teaching approaches, supportive environments, and conducive policies~\cite{bhattacharjee2020older}. Ensuring that older adults can navigate the digital world with confidence and autonomy is crucial for their full participation in an increasingly digital society which was the main implication of our study. 
Highlighting the fears and barriers older adults face in acquiring digital skills is crucial. However, this study aims to address these issues by consulting digital educators and providing practical suggestions for improvement.

\subsection{Older Adults Learning Digital Skills in Ireland}
The integration of technology into the lives of older adults in Ireland presents both opportunities and challenges. Hogan highlights that while computer use among older adults is increasing due to courses from Active Retired Associations~\footnote{Active Retired Association Ireland: \url{https://activeirl.ie/ari-mission/}} (ARA), significant technophobia persists~\cite{hogan2006technophobia}. These courses improve attitudes towards computers and reduce technophobia but do not significantly lower anxiety, indicating the need for more personalised teaching approaches~\cite{hogan2006technophobia}. Digital engagement among older adults in Ireland and Finland varies with socioeconomic status and education~\cite{pirhonen2020these}. While digitalisation offers benefits, it also leads to feelings of disconnection and loss of control, particularly among Finnish older adults. The study stresses that the digital divide is more about resource availability than ability, emphasising the need for inclusive policies and targeted support~\cite{pirhonen2020these}. Support for older adults in digital education in Ireland is delivered through a combination of family support, informal peer learning, and digital literacy initiatives via structured education courses. Flynn explores the preference for informal peer learning among older adults in Ireland, which is more effective than formal courses~\cite{flynn2024keeping}. This aligns with Marsick and Watkins~\cite{marsick2015informal} on informal learning and Topping~\cite{topping2013trends} on peer learning. Flynn also notes that structured government-provided adult education is primarily personalised for beginners and low-tech users~\cite{flynn2024keeping}. 

Community-based programmes such as `Getting Started', provided by the NGO Age Action Ireland, have empowered over 46,000 older adults to develop essential digital skills, bridging the digital divide and enhancing their confidence in technology usage. The `Hi Digital' initiative, led by Vodafone and ALONE Ireland, leverages peer-led training to enhance practical digital skills among older adults. Additionally, a 10-year adult literacy strategy launched by SOLAS in 2021 aims to reduce the percentage of adults lacking basic digital skills from 47\% to 20\%~\cite{hi-digital}.  

\subsection{Engaging Digital Supporters}

Although much research on digital literacy in older adults focuses on interviewing the older adults themselves, there is limited research involving the educators directly. Tomczyk et al. examine digital inclusion from the perspective of the educator, highlighting challenges such as varying skill levels among learners, resistance to technology, and limited resources~\cite{tomczyk2022digital}. They emphasise that effective support includes personalised instruction, patience, and practical applications of technology. Tailored approaches and adequate support are crucial for enhancing digital inclusion among older adults. Lobuono et al. explore methods of teaching older adults used by student mentors, identifying eight different modalities: observation and listening, writing down information or creating visual aids, explaining and simplifying material, using repetition and review, and a hands-on approach~\cite{lobuono2020teaching}. These personalised, hands-on methods are adapted to suit individual learning needs, underscoring the importance of flexibility in teaching styles. Arthanat et al. emphasise the role of motivation, the trainer-trainee relationship, patience, self-reliance, and mutual value as key strategies for effectively implementing ICT training for older adults~\cite{arthanat2019multi}. Similarly, Schirmer et al. advocate for ICT training that considers the life experiences of older participants. They suggest that aligning training with the needs, values, and desires of older adults effectively integrates new skills into their lives~\cite{schirmer2022digital}.

Geerts et al. investigate how digital instructors support the acquisition of digital skills in later life, examining their role in promoting digital inclusion among older adults~\cite{geerts2023bridging}. Their study highlights the importance of personalised instruction tailored to older learners' needs and calls for more recognition and support for digital instructors. They advocate for policies that promote sustained, individualised digital education for older adults. Gates et al. conduct a systematic narrative review to investigate the implementation and delivery of digital skills programmes for middle-aged and older adults, examining the integration of adult learning theories such as geragogy and critical geragogy~\cite{gates2022role}. They identify three recurring themes: negative perceptions of ageing, the importance of the learning environment, and the value of technology in these programmes. Vercruyssen et al. examine the complexities surrounding digital literacy for older adults from the perspectives of digital skills instructors. Interviewing 26 digital instructors in Belgium, they find that `basic' digital literacy for older adults encompasses more than just technical skills. It involves addressing emotional, cognitive, and social dimensions, which has significant implications for teaching practices and policy development~\cite{vercruyssen2023basic}. Along these lines, Chiu et al. present findings from interviews with internet technology instructors at a senior learning centre, identifying key strategies for effective education: simple instructions, hands-on practice, personalised support, and a patient, encouraging environment~\cite{chiu2019help}. Instructors highlighted the importance of relating technology to seniors' daily lives and boosting their confidence with positive reinforcement. This study underscores the effectiveness of tailored teaching methods in enhancing older adults' technological skills and learning experiences.  These studies collectively highlight the importance of flexible, patient, and supportive teaching methods tailored to the unique needs of older adults. Our study builds on these insights, emphasising the necessity for sustainable and inclusive digital education practices that empower older adults to navigate the digital landscape confidently and autonomously.

\section{Methodology}
This section provides an overview of the methodology employed in our study, detailing the processes involved in sourcing digital hubs, participant recruitment, participant demographics, procedure, and data collection and analysis in Ireland.

\subsection{Sourcing Digital Hubs}
Prior to the recruitment phase, we conducted an extensive internet search to identify all digital learning hubs in Ireland. This involved a thorough review of existing research reports, online directories, and leveraging the research team's extensive knowledge of existing resources. We compiled a comprehensive database of contact numbers for digital hubs and created a list of age-friendly libraries. This database served as a crucial resource for the subsequent participant recruitment process, ensuring that we had a broad and representative sample of digital learning providers across Ireland. The sourcing of the digital hubs was done by the primary author of this paper.

\subsection{Participant Recruitment}
The initial aim was to interview volunteers providing technical support to older adults. However, the scope of recruitment expanded to include digital educators and anyone in the Republic of Ireland providing educational services on digital skills to adults aged $65$ and older. We contacted $66$ organisations known to provide digital skills services to older adults. These included libraries, education and training boards (ETBs), older adults service NGOs, adult literacy groups, professional training colleges, community centres, and county councils. Organisation leaders were briefed on the study and provided with a detailed study brief, a text pitch, and an offer of a \euro40 `One-4-All' voucher as compensation for participation. We employed a combination of direct recruitment and snowball sampling. Potential participants identified by organisation leaders were invited to participate, and participants were encouraged to spread the word within their communities and group chats after being interviewed, which significantly accelerated recruitment. 

\subsection{Participant Demographics}
\begin{figure}
    \begin{subfigure}{0.45\textwidth}
        \includegraphics[width=\textwidth]{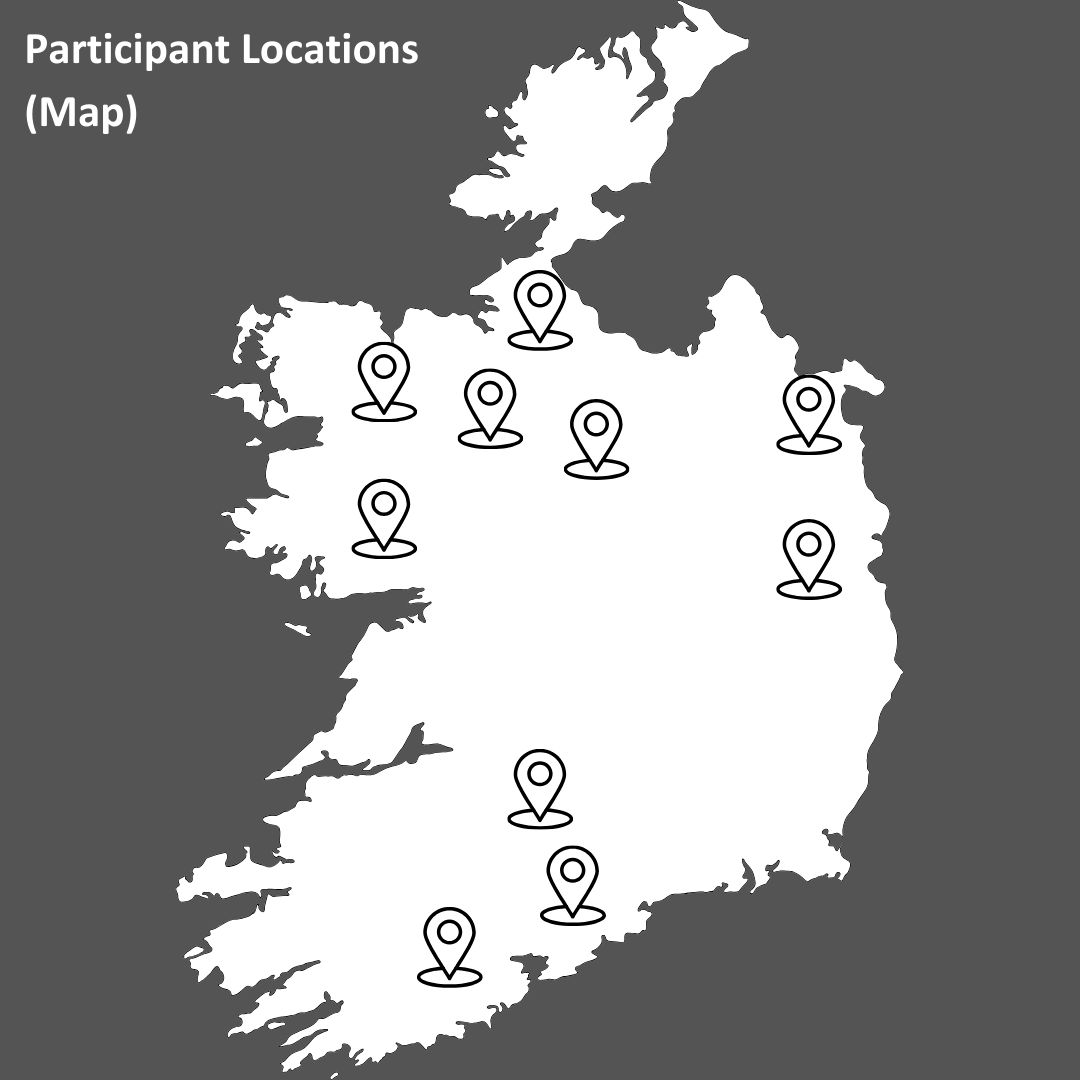}
        \caption{Distribution of study participants across the Republic of Ireland.}
        \Description{}
        \label{fig:map}
    \end{subfigure}
\hfill
\begin{subfigure}{0.45\textwidth}
    \centering
    \includegraphics[width=\textwidth]{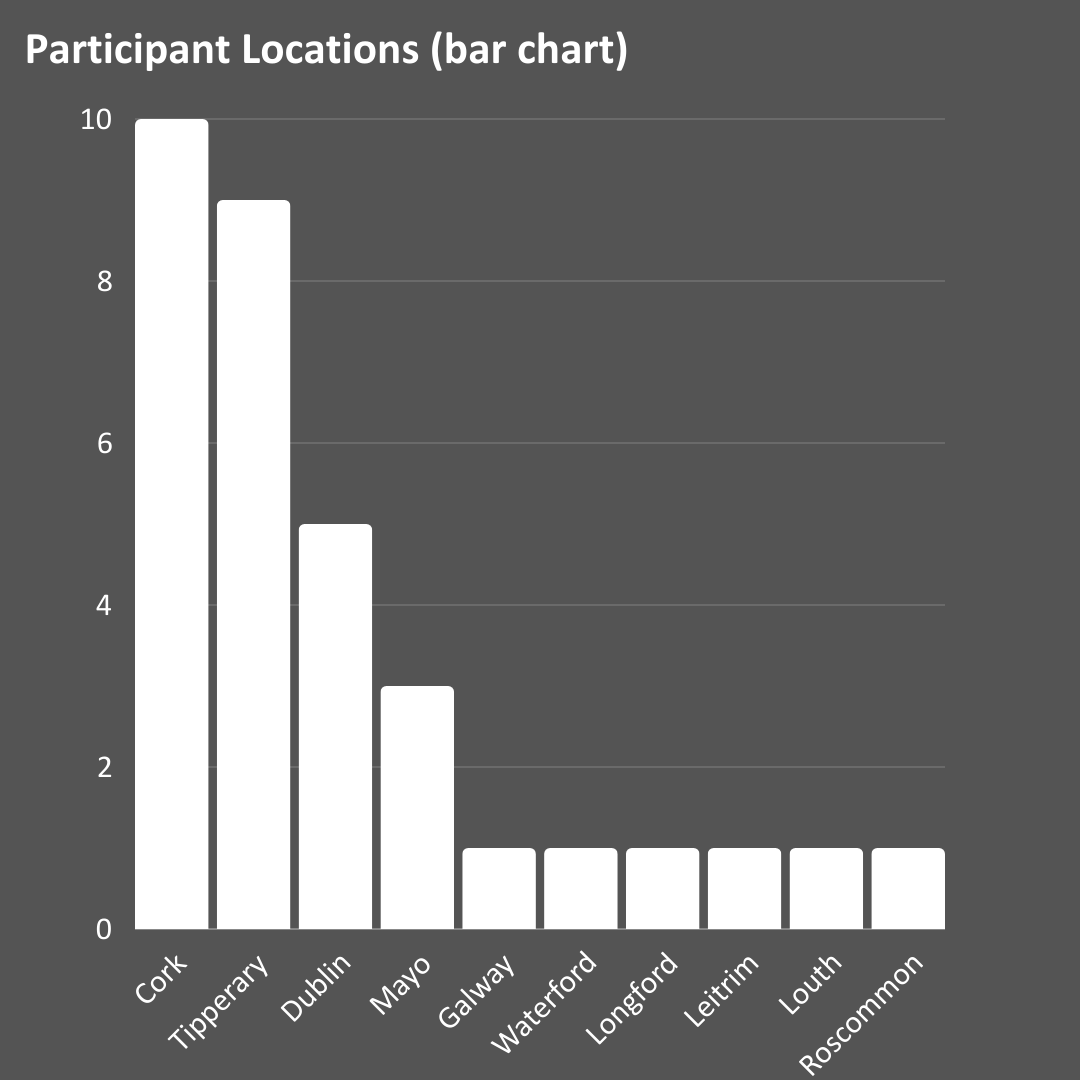}
    \caption{Bar chart showing the counties participants in our study are located in.}
    \Description{}
    \label{fig:barchart}
    \end{subfigure}
    \caption{Participant Locations.}
    \Description{Two images show the distribution of participants in our study. The first image shows that there is a good regional spread across north south east and west of Ireland. The second figure shows the number of participants from each county in Ireland. There were 10 participants from Cork, 8 from Tipperary, 5 from Dublin, 3 from Mayo and 1 from Galway, Waterford, Longford, Leitrim, Louth and Roscommon.}
    \label{fig:map-bar}
\end{figure}

Our study recruited $41$ participants, comprising $12$ males and $29$ females. After accounting for moderate attrition, $34$ participants ($9$ males, $25$ females) completed semi-structured interviews. Participants represented a diverse range of both rural and urban regions across the Republic of Ireland, including counties Dublin, Cork, Tipperary, Galway, Leitrim, Louth, Roscommon, Longford, Mayo, and Waterford. This geographical diversity is illustrated in Figure~\ref{fig:map-bar}. As we see there was a total of 9 male participants (26\%) and 25 female (74\%). This is about on par with Ireland's gender breakdown within the education sector according to Ireland's Central Statistics Office (CSO) report on gendered demographics in Ireland in 2019. In 2017 the CSO found second level classroom teachers to be 29.7\% male and 70.3\% female, and graduates specialising in the education field were found to be 25.59\% male and 74.41\% female~\cite{cso2019womenmen}. 

Participants were affiliated with various types of organisations. Ten participants volunteered for older adult services charities, six were paid employees of non-governmental adult literacy organisations, one worked for a for-profit adult education college, one was employed by a corporation providing drop-in services, and $17$ were government employees working with ETBs. The experience levels of these digital educators in providing digital skills services to older adults ranged from three months to over $30$ years, with an average of $7.05$ years ($SD = 6.92$). 

\subsection{Study Protocol}
To ensure participants were well-informed about the study, we sent a short description via email, along with a longer project summary in PDF format. Interviews were 
conducted over Microsoft Teams, during which, 
we read an 
informed consent statement to participants, seeking verbal consent and noting their responses. With consent obtained, we recorded the meeting audio to generate transcripts. The interview protocol comprised questions, focusing on participants' tenure as digital educators, their locations, the strengths of their programmes, the barriers they and their learners encountered (to their knowledge), and their desired future changes in digital education for older adults. At the conclusion of the interviews, participants were asked for their postal addresses to receive the promised One-4-All voucher. This study was approved by the ethical review board of the organisation.

\subsection{Data Analysis}

The process of interviewing participants and collecting data spanned two months.
Transcripts were automatically generated from the recorded audio and subsequently cleaned by the research team.

Once the coding was complete, we moved on to thematic analysis. In this study, we adopted reflexive thematic analysis to examine the interview data, following the six-phase process outlined by Braun and Clarke~\citep{Braun2019,braun2012thematic}. In line with the reflexive approach, we engaged in an iterative and reflective process throughout, prioritizing flexibility and the development of researcher insights at each stage. This process began with familiarisation, during which the researcher immersed themselves in the data, followed by coding to capture key patterns and meanings. We then generated initial themes, which were continuously developed and reviewed, allowing themes to evolve organically as our understanding deepened. In the refining stage, we worked to clearly define and name each theme, emphasizing the nuanced interpretations and insights drawn from the data. Finally, we completed the process by writing up, ensuring that the themes provided a cohesive narrative addressing the research questions.

Through this reflexive approach, we identified $39$ initial codes, which were distilled into $16$ overarching themes that highlighted the benefits and barriers faced by older adult learners and their educators. Throughout the results, we report on representative quotations to illustrate each theme. These quotations provide examples of the participants' experiences and perspectives, grounding our results in the interpretations conveyed in the data. The analysis also revealed a variety of digital education programs offered by different educators, though course frequency and class sizes varied among participants.

\section{Results}
In conducting our thematic analysis, we extracted 39 codes and based on these codes, we found 16 themes. These themes revealed barriers and solutions for digital education programmes geared toward older adults. In this section methods for expanding digital skills programmes, supporting digital educators, and removing systemic roadblocks for older adults are discussed. 

\subsection{Maximising Design and Implementation of Digital Skills Programmes }

\subsubsection{Learner-led }
From talking with digital educators we found it to be crucial for digital education to be tailored to older adults' specific wants and needs. Of all themes the learner-led theme occurred most frequently in the interviews (23 out of 34 interviews) and we found several dimensions of benefits.  Schirmer et. al note the benefits of taking into account personal experiences of older students, and ``contend that successful instructors take the lifeworld of the participants into account, adjust the content and the examples geared toward interests and practical relevance for the audience''\cite{schirmer2022digital}. We found older adults' learning desires and needs may differ from other populations and giving them a sense of agency in the direction of the learning can lead them to feel more invested. One digital educator observed that the tailor-made approach can ingratiate older adults to the programme.

\begin{quote}``And that's why I think they like our event because we don't come and assume, `this is what you need to know about.' We ask them what do you need? What do you need to know? What are you struggling with? Tell me how can I help you?'' (P34) 
\end{quote}

We found it to be advantageous for digital educators in structured learning environments to solicit the feedback of older adults on day one of the class or workshop, as this could increase buy-in. We found that it may also reduce attrition by preventing older adults from feeling overwhelmed on the first day of the course. Eighteen out of 34 interviewees sought input from older adults on the first session; this significant finding could inform the development of digital education programmes.

\begin{quote}
    
``When I start a group, any group, my first day I ask them what would they like to learn.'' (P25)   \end{quote}

We discovered powerful impact not only from digital educators seeking group feedback, but from finding out older adults' individual interests and using that as a guide. Some digital educators use older adults' specific topical interests as a way to drive learning forward and secure the investment of learners. Digital educators successfully equip older adults to pursue their unique interests through technology, with examples such as using Spotify to access music while travelling, looking up an ABBA concert on YouTube, and learning web browsing through looking up sport results.

\begin{quote}``I try to ask them questions, find out what they're most interested in and what's going on in their lives and then try and build a solution around that and find apps that I think might be interesting to them.'' (P17)
\end{quote}

In Ireland's digital skills education landscape, some types of services are more flexible than others; for example, courses taught at education and training boards may have a 
curriculum as a guide. NGO's and drop-in services can be more open-ended in format. While we did not find structured curriculum to be a disadvantage, flexibility was a boon for tailoring the programme. Flynn also found flexibility to be an advantage with regard to peer learning amongst older adults~\cite{flynn2024keeping}.

\begin{quote}``That was one of the major advantages I think, that we've no set curriculum and the older person is in the driving seat.'' (P1)\end{quote}

\subsubsection{Step-by-Step Instruction}
Step-by-step instruction appeared to be key in digital skills teaching settings. Noted also by Ahmad et al. \cite{ahmad2022effectiveness}. When learning new technology, digital educators said that older adults benefit from piece-by-piece instructions at a reasonable pace, combined with repetition where necessary, which Geerts et al also stipulate \cite{geerts2023bridging}. Even older adults with low technology confidence can quickly pick up on new concepts with the right kind of instruction.

\begin{quote}``Because during that smartphone course, it's amazing to see how if you just give them the right steps and just do it at a certain pace and do it frequently enough each week, they're well able to do it; they're well able to take on the new information.'' (P27) 
\end{quote}

Participants in our study felt they would benefit from extra course materials, such as booklets and printable PDF's, that contain step-by-step instructions. However, any take-home materials are likely to need supplementary in-person instruction, because as digital interfaces are updated, step-by-step instructions may become obsolete. This reaffirms the importance of digital educators in empowering older adults to be connected and independent.

\subsubsection{Conceptual Understanding} 
In eight interviews digital educators reported that older adults are often interested in a conceptual understanding of technology; they wish to learn more than the steps to complete the task, also desiring technological context. Rather than merely knowing `how to use the Internet' they may wish to know `what is the Internet' or `how does it work'. If digital educators are teaching the use of a web browser, they may be fielding questions about what other types of web browsers there are, why there are so many types of the same functional programmes, and what is going on the back end of a Cloud service. Some digital educators may not know how to answer these questions. Learning how to explain the `nuts and bolts' – for example the difference between hardware, software applications, and system software– may be a helpful type of upskilling for digital educators to cater to older adults' learning needs.

\begin{quote}``Even people [digital educators] who are tech-savvy may not necessarily have that conceptual understanding of the physical things that we can touch, the operating system, software, application and apps.'' (P19)
\end{quote}

Older adults, not being 'digital natives,' may desire a more fundamental understanding of technologies rather than taking them for granted.
Indeed, past digital skills research has pointed to the `critical adopters' trait found in older people, showing that the friction they experience with new technology can be a source of insight on how we build and teach technology~\cite{barros2021circumspect}.
Some of our digital educators indicated this conceptual approach is an alternative way of looking at technology:

\begin{quote}``With small children when they pick up a phone, or teenagers, they take it on face value. If they're shown something once, they say, OK, that's the way it works. They don't need to know why it works. For my generation, I suppose it's because it's so new and it's so far outside our our upbringing and everything else that we we don't really know why it works. And we wonder why it works; there is that block there. '' (P26)
\end{quote}

\subsubsection{Social Benefits }
Our study revealed that digital educators observe that the social benefits of attending digital skills courses is a draw for older adults. They reported it is also helpful in their learning. According to digital educators, many Irish older adults attending digital skills courses enjoy utilising the time for forming or growing connections with other learners or with digital educators.

The first day of digital skills courses brings risk of attrition, but we found that the social aspects of attending digital skills classes had a positive impact on continued learner attendance. Older adults seem to benefit from realising on the first day that courses will be informal and will have space for `little chats' (P12). Another benefit of social time in digital skills courses is a break from information overload.

\begin{quote}
``They get tired learning new stuff and if you try and throw too much at them in two hours, it is too much. They also like to come for the social side.'' (P17)
\end{quote}

Digital educators indicated that the social aspect of the learning setting can sometimes reduce shame for older adult learners. Learning that other older adults are also struggling may be a helpful part of the programme for many older adults. 

\begin{quote}
``I find the biggest single aspect of the programme that is a help, is when people realise there are other people out there that feel like them.'' (P33) \end{quote}

One social challenge that exists in some digital skills courses for older adults is that bonds are formed between digital educator and learner and in many cases digital educators are expected to cut ties for safeguarding reasons. Many Irish older adults are socially isolated \cite{tilda-lonely} and it can be challenging for both parties to break that bond. One NGO volunteer recommended a part-time digital skills drop-in centre after the programme's completion, so that older adults could loosely maintain the social ties with volunteers that had been built up over the weeks. We believe Ireland would benefit in general from additional drop-in digital skills centres as one-to-one time is in high demand within structured courses.

\subsection{Barriers Older Adults Face in Learning Digital Skills}

\subsubsection{Fear}
We found that digital educators reported fear to be the most prominent barrier for older adults in learning digital skills and using technology. Twenty-five out of 34 interviews mentioned fear as a barrier to learning and use of technology. Twenty-six interviews mentioned the prevalence of scams, and 12 mentioned lack of confidence as a challenge for older adults. Some older adults' fears seemed attuned to the number of cyber threats present in the greater digital landscape, but other fears may be overly cautious.  

\begin{quote}``So they'd have all sorts of fears of, `I'm going to press the wrong button and I'm going to kill the phone' or `I'm going to press the wrong button and I'm going to delete my whole phone book' or `If I lose somebody's number, I can never get it back again.' '' (P21)\end{quote}

A sense of empowerment with technology, gained through digital skills education, can help older adults feel less afraid. The role of digital educators is not just to reassure older adults about their fears, but to guide them through exploring their technology until older adults feel more confident. Fifteen out of the 34 digital educators we interviewed described older adults feeling empowered due to learning.

\begin{quote}``And I do know that there's a lot of fear there. And if you can equip people with information, sometimes that can be a really good tool in their arsenal to help them feel more confident, more empowered.'' (P33) \end{quote}

\subsubsection{Inaccessible Language }
Digital educators indicated that a significant barrier for older adults learning digital skills is inaccessible language (13 out of 34 interviews). Dense verbiage, walls of text, complex words, niche terminology, and small text are factors that make it more difficult for older adults to advance their skills. Digital educators emphasise the importance of simple language, both in teaching and in educational materials, as technical jargon runs the risk of making older adults feel overwhelmed. 

The `learner-led' or tailoring theme also appeared entwined with the idea of explaining concepts simply, where one digital educator stated the importance of tailoring explanations to the right level of complexity.

\begin{quote}``I've been teaching for years; I know I need to tailor my language exactly to the audience that I'm speaking to. You need to just give a certain amount of information at a certain time or tailor the complexity to what the person knows coming into the situation.'' (P21).\end{quote}

Previous studies support the idea that technical jargon reduces older adults' confidence in being able to use technology~\cite{marston2019older,neves2021digital}. It will be crucial going forward to have simple language in digital skills materials and programmes geared toward older adults. 

\subsubsection{Outdated Devices}
We found that digital educators report outdated devices as a barrier to older adults learning technology and gaining digital independence.  Six interviews indicated that older adults often present to digital skills programmes with outdated devices. This observation aligns with findings from Tomczyk et al.~\cite{tomczyk2023barriers}. These devices include laptops too old to run programmes on, phones too old to keep secure, and devices in poor condition. Older adults' devices are often older devices gifted from younger family members or not upgraded due to fixed income. This can prove to be a challenge for digital educators in their teaching as devices incompatible with newer programmes may be either not usable or not safe.

\begin{quote}
``If they bring in an old machine or an old phone that won't do updates and won't do virus checking, I'd say, `You can't use that anymore. That's like leaving your house open.' '' (P11)\end{quote}

\subsubsection{Lack of Family Support in Digital Education}
Digital educators encounter another type of barrier for older adults: family tech caregivers often refuse to teach digital skills but will instead complete the task on behalf of the older adult. Eleven of the 34 interviews with digital educators raised this issue of family being unwilling to teach. Digital educators told us this robs older adults of the autonomy to learn. Additionally, a younger person taking a device from an older adult and completing the task for them may create a sense of shame around an older adult's skill level.

 \begin{quote}
 ``If they're asking for help from younger people in particular, the young person, rather than actually showing them, will typically just say `God' and just do it for them really, really quickly. And that almost makes them worse. So that's the biggest barrier I think for people; it can create so much shame.''(P16)
 \end{quote}

These limitations with family caregivers highlight the importance of expanding digital skills programmes for older people, but as we will discuss later these programmes in Ireland can often be underfunded or short on time resources. Family caregivers are not a replacement for digital skills services; %O'Keeffe et al. (2024) 
our preliminary study found that 25\% of older adults have no sense of family support when it comes to cyber safety.  Morrison et al. found that relying on family and friends resulted in less learning and confidence \cite{morrison2021older}. Geerts et al. also found this to be the case comparing family `warm' with formal `cold' education \cite{geerts2023bridging}.
Reliance upon friends and family produced more mixed results, with less learning and poorer long-term effects upon user confidence \cite{morrison2021older}.

Consistent with Rosales and Blanche-T~\cite{rosales2022explicit}, digital educators in Ireland are informed that younger people show frustration and a lack of patience when older adults seek help in completing an online task. Due to this barrier, some older adult learners reportedly avoid asking family altogether.
 
\begin{quote}
``They would say, `I can't ask my daughter or my son this because they'll just get mad at me.' '' (P33) 
\end{quote}

\subsection{Systemic Issues Potentially Worsening the Digital Divide}

\subsubsection{Time Constraints }
Time constraints are a major barrier in the world of digital skills education. When digital educators were asked what their main barriers were, time constraints were prominent; the word `time' came up frequently, mentioned in 22 instances within the data. Thirteen of 34 interviews showed that digital educators and older adults would benefit from additional time allotted within the digital skills programmes. This included class time, paid preparation time, length of programmes, and frequency of services. Most prominent, however, was the need for additional one-to-one time, where educators could work directly with an older adult on their specific concern either in or outside the classroom. The benefit of allowing enough time in a session to teach a digital skill was also highlighted in Gates et al.'s systematic review exploring the implementation and delivery of digital skills programs for older adults~\cite{gates2022role}. 
Digital educators called for more frequency in drop-in sessions.

\begin{quote}
    
``If we're looking at addressing the digital gap specifically, I think there should be more frequency [of digital education sessions], or even a standing one-stop-shop for them; they know that they can go there, they'll get the help every day whenever needed. Once per quarter is not enough to address anything.'' (P34) 
\end{quote}

Additional one-to-one time was found to be crucial, as class size and frequency limits do not always allow for enough direct tutoring to adequately support older adults. Drop-in centres may be a good solution to this, but regardless of service delivery type, more hours need to be allotted.

\begin{quote}
``My barriers in particular as a tutor are, you might only have six weeks to teach them. And we have to have class sizes of six or more. So six people getting to know a technology that they've never known before within six weeks makes people feel like they're not getting the attention they need. The most ideal situation I think for people would be one-to-one sessions with no dictated timeline.'' (P15)
\end{quote} 

\subsubsection{Lack of Funding } 

There is a dire need for increased funding for digital skills programmes. Seven digital educators cited lack of funding as a barrier for digital skills programmes for older adults. Funding limitations affect the length, class duration, scope, and size of digital skills programmes. 

\begin{quote}
     
``It always comes down to funding. We have a waiting list for our courses. We could probably run 10 times the amount of courses, but there isn't the funding available.'' (P30) 
\end{quote}

Other consequences of limited funding are lack of advertising, lack of compensation for continuing professional development and unpaid preparation hours in the case of paid tutors, and a heavy reliance on volunteer labour. As a reminder, 10 out of 34 of our digital educator participants were unpaid. One volunteer recommended incentives for volunteers to help with retention in areas with smaller teams. 

\begin{quote}
    
``So I'd be saying for the volunteers of the future, just to put in some kind of an incentive. And also, maybe for more of a marketing campaign to sell what is on offer. Because it's absolutely amazing when it works and should just have that bit more funding from the government bodies.'' (P7) 
\end{quote}

\subsubsection{Rural Internet Issues }
Interviews with digital educators (8 out of 34) revealed that in rural areas, many older adults cannot access broadband services at all. Some technical support services examined in this study are delivered remotely, making lack of internet access a significant challenge. Digital educators indicated that many older adults in rural areas lack the ability to practice computer skills at home and do not understand how the internet works, unsure about using their phones due to fears about a high data use bill.  

\begin{quote}
    
``A lot of them would be in an area where you have bad broadband connection, where they're not able to connect to the Internet.'' (P5) 
\end{quote}

The Irish National Broadband plan~\cite{NBI} has been heavily criticised in Irish media for its delay of broadband coverage for rural Ireland~\cite{Irish-examiner}. In 2021 the Eurobarometer research findings found that Ireland was one of the worst countries in addressing the rural broadband gap~\cite{Irish-Indo-Broadband,EU-Barometer}.  
According to the Digital Inclusion in Ireland report, in 2019, 92\% of households in Dublin had a fixed broadband connection, compared to 69\% of households in the Midlands~\cite{NESC}. With 25\% of those surveyed in rural areas reporting daily problems with the quality of mobile-phone coverage~\cite{NESC}. 

\subsubsection{Lack of Promotion}
Digital educators reported that the lack of promotion and advertising for digital skills programmes in Ireland proves a significant barrier to reaching older adults. Nine out of 34 digital educators underscored the necessity of spreading the word about existing digital skills programmes. It seemed that many older adults do not know where in their community to go for help, especially where there is a lack of family support. Reaching digital skills learners poses a unique challenge as free social media advertising may not be likely to reach older adults; social media use is a learning target for many learners in these very courses. Other types of advertising may be more effective, but funding re-emerges as a barrier in that case.

\begin{quote}
    
``There's no point putting on a Facebook post. We're doing a course for seniors. If the seniors are not on Facebook, then that's the issue. Who's got the funding to do a mail drop?'' (P30) 
\end{quote}

Digital educators recommended going where older adults can be found: libraries, Men's Sheds, and grocery stores were suggested. One digital educator recommended church newsletters. Another drew out the fact that isolated older adults may be going to libraries to feel more connected, and could encounter physical advertising there. Radio would be an advantageous approach for reaching older adults about digital skills programme offerings.

\begin{quote}
    
``I definitely think the local radio is the best way to reach (them), because the older generation would always have the radio on in the house.'' (P12) 
\end{quote}

\subsubsection{Lack of Public Transport to Courses}
Older adults may need increased access to transport in order to access digital skills education services. Limited public transport was found as a barrier in five interviews, where digital educators described limited transport connections to courses and drop-ins as a source of digital isolation for older adults. When asked what would be most helpful in removing barriers, P7 answered, 

\begin{quote}
    ``Gosh, I suppose more rural transport, but sure, everyone is saying that.'' (P7)
\end{quote}

One NGO employee, P24, cited the helpfulness of remote digital skills services in light of transport issues; however, only one out of 11 digital hubs researched in our interviews offered remote services.

\begin{quote}
    
``The remote (service) is great for people who don't have access to transportation or have health issues or different reasons why they can't go to a centre.'' (P24)
\end{quote}

Ireland is referenced as one of the European countries where over 60\% of the population consistently reports that reaching places by public transport is difficult or impossible~\cite{EU-Barometer}. A recent study revealed that access to public transport is a major challenge in rural areas, with 47\% of respondents lacking access to public transport, including local link services ~\cite{transport-justice}. 
One interviewee recommended a shuttle for helping older adults get to digital skills programmes. Four additional interviews revealed that older adults must adapt and learn to book train station parking and bus tickets online as in-person bookings disappear. Without digital skills training programmes, digital educators note these changes can leave older adults unable to travel to attend appointments or see loved ones. 

Lack of digital skills was likened to a new type of illiteracy, especially with advance bookings increasingly being required for bus travel.

\begin{quote}
    
``If you can't use digital skills, it's like not being able to read and write. If you want to get on the bus, you have to book your ticket beforehand.'' (P22)  
\end{quote}

The many systemic issues which prevent older adults from accessing digital skills education services point to a greater problem within Irish society: the deprioritisation of social programmes and subsequently of vulnerable groups. The lack of focus on systemic structures to support older adults is perhaps a cornerstone in a greater discussion about the purpose of society.

\subsection{Solutions to Support Digital Educators}

\subsubsection{Upskilling }
We found a demand among digital educators for upskilling, with a special urgency around cyber security. Twelve out of 34 digital educators indicated they would benefit from upskilling. 
When asked what the barriers are to helping older adults, one digital educator (P10) noted; ``Number one, my own training. My own technical skills. That's definitely an issue for me and I can only go so far.''  

Some wished to stay up to date and advance their existing knowledge to keep up with the times; others felt generally under-equipped to field the incoming questions from older adults. As mentioned in the introduction, the interviewees were volunteers and workers in NGO's and education \& training boards (ETB's). Between NGO and ETB digital educators, ETB digital educators were more likely to feel confident delivering modules. However, the issue of unpaid or costly continuing education was raised by ETB tutors. 

When asked what would be helpful in removing barriers for digital educators, a digital educator (P31) shared, ``Even an online course to train us.'' 
Another digital educator admitted their team is often unsure about the answers.

\begin{quote}
    
``We're kind of guessing half the time.'' (P3) 
\end{quote}

Digital educators are, for many older adults, the first line of defence against a confusing and risky online landscape.  Though digital educators are skilled in teaching and communication, our findings suggest that their knowledge of cyber safety may not be strong enough to effectively support services for older adults. An emphasis on developing free upskilling materials and training, especially advised by tailored cyber safety advice, could make a meaningful impact for them.

\subsubsection{Relevant Course Materials }

Digital educators and their learners would greatly benefit from the development of relevant course materials. Nearly half the interviews, 15 out of 34, revealed desires for up-to-date cyber safety and digital skills resources which digital educators could use in their courses. Digital educators were interested in PDF resources, printed leaflets and booklets, websites, and e-learning courses. They cited the helpfulness of online tools such as FraudSmart and pre-existing paper resources from NGO's. Digital educators emphasised the importance of materials being simple and relevant. 

\begin{quote}
    
``Cyber security is huge. I would like even the likes of a small booklet just on the pitfalls. 
 What to look out for. What are the main issues in cyber security?'' (P25) 
\end{quote}

\subsubsection{Online Banking Training Platform }

A highly niche sticking point came up in five interviews: digital educators lack a tool to teach online banking. Safeguarding issues are posed in both directions when a live banking portal is used for training purposes; digital educators are not comfortable showing their own bank details or asking older adults to share theirs. 

\begin{quote}
    
``The banking one is difficult because it's one that comes up for people a lot. But you can't be using people's bank details.'' (P23) 
\end{quote}

Two digital educators recommended a dummy banking platform, similar to what some banks use when training their own employees on online banking. This would simply need to be kept up to date with the bank's current interface, and even one major bank offering this could prove helpful for expanding older adults' access to online banking education. 

\begin{quote}

``So we can't — I've never really got into teaching people banking online. I don't want to see their bank account. I don't want them to see mine. So I've always promoted that banks and the car tax and gov.ie\footnote{A central Irish website for government services and information.}, they should all give us dummy logins to go in and log in as a learner.'' (P11)
\end{quote}

Furthermore, with banks pushing for online services ~\cite{keogh2021growth}and reducing customer services teams~\cite{rajput2019impact}, we suggest commercial banks have a responsibility to set up supports so that older people are not forgotten and left financially vulnerable.
Latulipe et al. emphasise the crucial role of unofficial proxies (family members, friends, or other community members) in aiding older adults with their banking needs. They note that these close associates often know the online banking credentials of the older adults they assist. Latulipe et al. also calls for the design of online banking systems to better acknowledge the nuanced and evolving role of these helpers.~\cite{latulipe2022unofficial}.

\section{Discussion \& Implications}
The findings of this study reveal significant implications for the digital skills landscape, particularly in supporting older adults and empowering digital educators. Below, we outline practical, actionable insights drawn from our analysis and propose pathways for improvement in programme design, policy-making, and infrastructure development.

\subsection{Tailoring Digital Skills Programmes for Personalized Learning}
One of the clearest implications from our study is the need for more personalized digital skills programmes. Many older adults benefited most when their individual questions and concerns were addressed one-on-one by digital educators. This need for personalized support reflects a broader requirement for programmes to be flexible and adaptable to learners' individual needs, recognizing that older adults learn at different paces and often require additional time to understand new technological concepts. Participants frequently mentioned that extending contact hours between educators and learners—enabling step-by-step, tailored instruction—would greatly enhance the learning experience. One participant remarked, ``Older adults often require more time to grasp the `why' or `how' behind technology, beyond just following instructions'' (P5). Moving forward, digital skills programmes should prioritize flexibility and human-centered teaching methods, offering more opportunities for individualized learning sessions and interactive support. Additionally, integrating more interactive and user-friendly digital tools can enhance older adults' engagement and confidence in learning new technologies.

\subsection{Addressing Fear and Building Cybersecurity Confidence}
A recurring theme in the study was the pervasive fear that older adults have around technology, particularly in relation to cybersecurity. Concerns about online scams and fraud were widely expressed, with participants citing fear as a significant barrier to digital engagement. Digital educators play a key role in overcoming this fear by offering reassurance and hands-on guidance. To address these fears effectively, digital skills programmes should incorporate dedicated cybersecurity education. Educators would benefit from up-to-date, practical resources that they can share with learners, providing clear, step-by-step instructions on how to stay safe online. Additionally, interactive learning approaches, such as mock phishing exercises or secure, practice environments, can give older adults the experience they need to feel more confident about navigating the digital world safely. Ensuring that these resources are simple, accessible, and relevant to older adults' specific concerns is critical in building their confidence and reducing the fear that currently limits their digital participation.

\subsection{Bridging Infrastructure Gaps in Rural Areas}
Participants in the study identified significant systemic barriers, particularly in rural areas where access to broadband and reliable transport is limited. For older adults in these regions, the lack of basic infrastructure exacerbates the digital divide and leaves them isolated from the benefits of digital engagement.  To address these disparities, it is crucial for policymakers to prioritize the expansion of broadband services in rural areas. Public investment in infrastructure is essential to ensure that older adults in remote locations can access the same digital skills programmes and resources as those in urban centers. In the interim, community-driven solutions, such as mobile internet hubs or government-funded transportation services to digital learning centers, could provide temporary relief and improve accessibility for older adults who currently lack reliable internet. These initiatives will help reduce the isolation felt by many rural older adults and create more equitable access to digital skills education.

\subsection{Promoting Corporate Responsibility in Digital Support for Older Adults}
As industries continue to digitize their services, older adults are at risk of being excluded from essential services such as banking, healthcare, and retail. Participants highlighted the responsibility that corporations, particularly banks, have in supporting older adults through this transition. We propose that corporations develop and provide training versions of their digital platforms, offering older adults the opportunity to practice using these platforms in a risk-free environment. For example, ``dummy'' online banking platforms, which simulate real-world transactions without the risk of financial loss, could be integrated into digital skills classes, allowing older adults to gain confidence in managing their finances online. In addition to these training platforms, banks should ensure that older adults can still access human support, either through in-person services or dedicated hotlines staffed by individuals trained to assist older customers with digital tasks. These initiatives would greatly reduce the anxiety older adults feel around managing their finances digitally and ensure that they are not left behind as services become increasingly digital.

\subsection{Increasing Funding and Expanding Programme Reach}
A key finding from the study was the overwhelming lack of funding available for digital skills programmes. Digital educators often operate with limited resources and are unable to meet the growing demand for digital education services for older adults. This underfunding leads to a shortage of contact hours, unpaid volunteer work, and the inability to scale programmes to reach more learners. To address these challenges, we recommend a multifaceted approach to securing more sustainable funding. Public funding must be increased to support the expansion of digital skills programmes, allowing for more educator hours, broader geographic reach, and the development of new resources. In addition to public investment, private companies that benefit from digital transformation—such as telecommunications firms, banks, and tech companies—should be encouraged to contribute financially to digital inclusion efforts as part of their corporate social responsibility strategies. These companies could provide grants, sponsorships, or in-kind support, such as donating digital devices, software, or training materials to digital education programmes. Increasing funding in these areas is vital to meeting the growing demand for digital literacy among older adults.

\subsection{Empowering Digital Educators with Resources and Professional Development}
Digital educators expressed the need for more comprehensive and accessible resources to help them effectively teach older adults. Many reported a lack of up-to-date materials, particularly in areas like cybersecurity, where the threat landscape is constantly evolving. In response, we propose the creation of centralized repositories of educational materials tailored specifically for digital educators working with older adults. These repositories should include printable guides, video tutorials, and interactive exercises that can be easily integrated into lessons. Additionally, ongoing professional development opportunities should be made available to digital educators, ensuring they remain equipped with the latest knowledge and skills needed to teach effectively. These opportunities could take the form of workshops, online certifications, or collaborative learning groups, and should be provided at no cost to the educators to encourage participation. Upskilling digital educators is a key step in strengthening the overall capacity of digital skills programmes, ensuring that they can meet the evolving needs of older adults.

\section{Limitations and Future Work}
Our study provides valuable insights into the role of digital educators in supporting older adults in Ireland. Interviewees often hesitated to critique their organisations, possibly due to confidentiality limits or cultural reasons, leading to initial claims of no barriers until we prompted for more detailed answers. A $2022$ linguistic analysis affirmed that indirect communication is common in Ireland, which we must consider in future research interviews~\cite{ronan2022directives}. We plan to replicate this work with a more geographically diverse sample and include older adults' perspectives for a holistic view. Additionally, we plan to conduct longitudinal studies which will help us track the adaptation of digital education over time. We also encountered libraries in Ireland where transition-year students (15-17 years old) provide technical support to older adults. Investigating this support from students could be a valuable research avenue. By exploring intergenerational learning models and tailoring programmes for specific subgroups, we can gain deeper insights. Finally, examining policy and institutional support is crucial for sustaining and scaling successful digital education initiatives. Addressing these areas will enhance support systems for older adults and their digital educators, ensuring they remain connected and empowered in the digital age.

\section{Conclusion}
Our research highlights the essential role digital education plays in empowering older adults to navigate the digital world with confidence. Interviews with digital educators revealed that programmes tailored to older adults' needs—offering personalized, step-by-step guidance—are crucial for effective learning. However, these programmes often face challenges, including limited time, funding, and access to clear, simple language. A key finding was the widespread fear of technology among older adults, which hampers their digital confidence. Overcoming this fear requires tangible and hands-on learning experiences, where digital tools are practised in a supportive, real-world context. This aligns with the importance of embodied interaction in learning, allowing older adults to engage directly with technology, building familiarity and confidence through repeated use. Systemic barriers such as poor broadband access and transportation further complicate digital inclusion, particularly in rural areas. These challenges underscore the need for embedding digital education within broader infrastructure improvements, ensuring that older adults have the necessary support to access services. Digital educators expressed a need for more accessible, up-to-date resources like booklets, PDFs, and interactive courses, particularly in the area of cyber safety. Simulated tools, such as an online banking platform, offer a practical way for older adults to engage safely with digital services, reducing anxiety and promoting long-term independence.

 \begin{acks}
We would like to thank the digital educators that participated in the interviews as well as each person within NGO's, libraries, and ETB's who spread the word about our study within their organisations.
This publication has emanated from research conducted with the financial support of the EU Commission Recovery and Resilience Facility under Research Ireland Our Tech Grant Number 22/NCF/OT/11212G. The research was supported in part by a Google Trust and Safety Research Award. Dr Murray is supported by Taighde Éireann – Research Ireland under Grant number 13/RC/2077\_P2 at CONNECT: the Research Ireland Centre for Future Networks.

\end{acks}

\bibliographystyle{unsrt}
\bibliography{Bibliography}

\begin{thebibliography}{10}

\bibitem{UN-report}
United Nations.
\newblock World population ageing, 2017.
\newblock \url{https://www.un.org/en/development/desa/population/publications/pdf/ageing/WPA2017_Report.pdf}.

\bibitem{WHO-active-ageing}
World~Health Organisation.
\newblock Active ageing a policy framework, 2002.
\newblock \url{https://iris.who.int/bitstream/handle/10665/67215/WH0?sequence=1\#page=19.10}.

\bibitem{millward2003grey}
Peter Millward.
\newblock The`grey digital divide': Perception, exclusion and barriers of access to the internet for older people.
\newblock {\em First monday}, 2003.

\bibitem{uk2019essential}
GOV UK.
\newblock Essential digital skills framework, 2019.

\bibitem{DESI}
European Commission.
\newblock The digital economy and society index (desi), 2022.
\newblock \url{https://digital-strategy.ec.europa.eu/en/policies/desi}.

\bibitem{centreforagingbetter}
Centre For~Aging Better.
\newblock The digital age: new approaches to supporting people in later life get online, 2018.
\newblock \url{https://ageing-better.org.uk/sites/default/files/2018-06/The-digital-age.pdf}.

\bibitem{gilmartin2018immigrant}
Mary Gilmartin and Jennifer Dagg.
\newblock Immigrant integration and settlement services in ireland, 2018.

\bibitem{davies2023healthcare}
Ben Davies.
\newblock Healthcare priorities: the “young” and the “old”.
\newblock {\em Cambridge Quarterly of Healthcare Ethics}, 32(2):174--185, 2023.

\bibitem{payne2022ageism}
Malcolm Payne.
\newblock Ageism, older people and covid-19.
\newblock In {\em Social work in health emergencies}, pages 201--215. Routledge, 2022.

\bibitem{minichiello2000perceptions}
Victor Minichiello, Jan Browne, and Hal Kendig.
\newblock Perceptions and consequences of ageism: views of older people.
\newblock {\em Ageing \& Society}, 20(3):253--278, 2000.

\bibitem{bradley1999history}
John Bradley.
\newblock The history of economic development in ireland, north and south.
\newblock In {\em Proceedings-British Academy}, volume~98, pages 35--68. Oxford University Press Inc., 1999.

\bibitem{wilson2023learning}
Gemma Wilson-Menzfeld, Jessica~Raven Gates, Mary Moreland, Helen Raw, and Amy Johnson.
\newblock Learning digital skills online: empowering older adults through one-to-one, online digital training provision.
\newblock {\em Frontiers in Psychology}, 14:1122277, 2023.

\bibitem{van2022zooming}
Jelle Van~Dijk.
\newblock Zooming in on embodied social sensemaking: Mapping the design space in the context of videoconferencing.
\newblock In {\em Proceedings of the Sixteenth International Conference on Tangible, Embedded, and Embodied Interaction}, pages 1--10, 2022.

\bibitem{harley2022together}
Daniel Harley, Stefan Grambart, Rodrigo Skazufka~Bergel, and Ali Mazalek.
\newblock Together alone: a tangible online narrative.
\newblock In {\em Proceedings of the Sixteenth International Conference on Tangible, Embedded, and Embodied Interaction}, pages 1--11, 2022.

\bibitem{pihlainen2023older}
Kaisa Pihlainen, Anja Ehlers, Rebekka Rohner, Katerina Cerna, Eija K{\"a}rn{\"a}, Moritz Hess, Lisa Hengl, Lotta Aavikko, Susanne Frewer-Graumann, Vera Gallistl, et~al.
\newblock Older adults’ reasons to participate in digital skills learning: An interdisciplinary, multiple case study from austria, finland, and germany.
\newblock {\em Studies in the Education of Adults}, 55(1):101--119, 2023.

\bibitem{cho2023internet}
Gawon Cho, Rebecca~A Betensky, and Virginia~W Chang.
\newblock Internet usage and the prospective risk of dementia: A population-based cohort study.
\newblock {\em Journal of the American Geriatrics Society}, 71(8):2419--2429, 2023.

\bibitem{Irish-Indo-gps}
Maeve McTaggart and Cian~Ó Broin.
\newblock Gp crisis investigation: Rural ireland worst hit as two-thirds unable to take on new patients, 2024.
\newblock \url{https://www.independent.ie/irish-news/gp-crisis-investigation-rural-ireland-worst-hit-as-two-thirds-unable-to-take-on-new-patients/a988706887.html}, Accessed: 25-06-2024.

\bibitem{hse-telehealth}
Health~Service Executive.
\newblock Hse telehealth roadmap 2024 - 2027, 2024.
\newblock \url{https://www.ehealthireland.ie/media/eo5nuaju/hse_telehealth_roadmap_full_report.pdf}.

\bibitem{mouratidis2021covid}
Kostas Mouratidis and Apostolos Papagiannakis.
\newblock Covid-19, internet, and mobility: The rise of telework, telehealth, e-learning, and e-shopping.
\newblock {\em Sustainable cities and society}, 74:103182, 2021.

\bibitem{chen2022workshops}
Amy Chen and Dominic Co.
\newblock Workshops in tei: Development, evaluation, exploration, and implementation: Development, evaluation, exploration, and implementation.
\newblock In {\em Proceedings of the Sixteenth International Conference on Tangible, Embedded, and Embodied Interaction}, pages 1--9, 2022.

\bibitem{zhang2024tablecanvas}
Yongxin Zhang, Charlotte~Mejlvang Guldb{\ae}k, Christian Fog~Dalsgaard Jensen, Nicolai~Brodersen Hansen, and Florian Echtler.
\newblock Tablecanvas: Remote open-ended play in physical-digital environments.
\newblock In {\em Proceedings of the Eighteenth International Conference on Tangible, Embedded, and Embodied Interaction}, pages 1--7, 2024.

\bibitem{weijdom2022performative}
Joris Weijdom.
\newblock Performative prototyping in collaborative mixed reality environments: an embodied design method for ideation and development in virtual reality.
\newblock In {\em Proceedings of the Sixteenth International Conference on Tangible, Embedded, and Embodied Interaction}, pages 1--13, 2022.

\bibitem{digital-age-action}
Age Action.
\newblock Digital inclusion and an ageing population.
\newblock \url{https://www.ageaction.ie/sites/default/files/digital_inclusion_and_an_ageing_population.pdf}, 2021.

\bibitem{fuchsberger2024remote}
Verena Fuchsberger and Lisa Hofer.
\newblock Remote, but tangible: Activities for grandparents and grandchildren across physical spaces.
\newblock In {\em Proceedings of the Eighteenth International Conference on Tangible, Embedded, and Embodied Interaction}, pages 1--7, 2024.

\bibitem{marani2021intensity}
Marco Marani, Gabriel~G Katul, William~K Pan, and Anthony~J Parolari.
\newblock Intensity and frequency of extreme novel epidemics.
\newblock {\em Proceedings of the National Academy of Sciences}, 118(35):e2105482118, 2021.

\bibitem{chen2021aging}
Yiyin Chen, Sabra~L Klein, Brian~T Garibaldi, Huifen Li, Cunjin Wu, Nicole~M Osevala, Taisheng Li, Joseph~B Margolick, Graham Pawelec, and Sean~X Leng.
\newblock Aging in covid-19: Vulnerability, immunity and intervention.
\newblock {\em Ageing research reviews}, 65:101205, 2021.

\bibitem{trinity-press}
The Irish Longitudinal~Study on~Ageing~(TILDA).
\newblock Research highlights increased loneliness in lock-down for the over-70s, 2020.
\newblock \url{https://tilda.tcd.ie/publications/reports/pdf/Report_Covid19SocialIsolation.pdf}.

\bibitem{finance-older}
Dr. John~A. Weafer and Dr.~Fergal Rhatigan.
\newblock A review of older people’s capacity to access financial services online and to independently conduct their own financial affairs, 2022.
\newblock \url{https://www.citizensinformationboard.ie/downloads/social_policy/Accessing_financial_services_online_report.pdf}.

\bibitem{thomas2023exploring}
Dain Thomas, Gobinda Chowdhury, and Ian Ruthven.
\newblock Exploring older people's challenges on online banking/finance systems: Early findings.
\newblock In {\em Proceedings of the 2023 Conference on Human Information Interaction and Retrieval}, pages 333--337, 2023.

\bibitem{mubarak2022elderly}
Farooq Mubarak and Reima Suomi.
\newblock Elderly forgotten? digital exclusion in the information age and the rising grey digital divide.
\newblock {\em INQUIRY: The Journal of Health Care Organization, Provision, and Financing}, 59:00469580221096272, 2022.

\bibitem{yu2023vulnerability}
Lei Yu, Gary Mottola, Christine~N Kieffer, Robert Mascio, Olivia Valdes, David~A Bennett, and Patricia~A Boyle.
\newblock Vulnerability of older adults to government impersonation scams.
\newblock {\em JAMA Network Open}, 6(9):e2335319--e2335319, 2023.

\bibitem{aarp-scam}
Jilenne Gunther.
\newblock The scope of elder financial exploitation: What it costs victims, 2023.
\newblock \url{https://www.aarp.org/content/dam/aarp/money/scams-and-fraud/2023/true-cost-elder-financial-exploitation.doi.10.26419-2Fppi.00194.001.pdf}.

\bibitem{shang2022psychology}
Yuxi Shang, Zhongxian Wu, Xiaoyu Du, Yanbin Jiang, Beibei Ma, and Meihong Chi.
\newblock The psychology of the internet fraud victimization of older adults: A systematic review.
\newblock {\em Frontiers in psychology}, 13:912242, 2022.

\bibitem{NCO}
Genevieve Waterman.
\newblock The top 5 financial scams targeting older adults, 2023.
\newblock \url{https://www.ncoa.org/article/top-5-financial-scams-targeting-older-adults}.

\bibitem{burnes2017prevalence}
David Burnes, Charles~R Henderson~Jr, Christine Sheppard, Rebecca Zhao, Karl Pillemer, and Mark~S Lachs.
\newblock Prevalence of financial fraud and scams among older adults in the united states: A systematic review and meta-analysis.
\newblock {\em American journal of public health}, 107(8):e13--e21, 2017.

\bibitem{MichiganUni}
University of~Michigan.
\newblock Experiences with scams among older adults, 2023.
\newblock \url{https://www.healthyagingpoll.org/reports-more/report/experiences-scams-among-older-adults}.

\bibitem{nicholson2019if}
James Nicholson, Lynne Coventry, and Pamela Briggs.
\newblock " if it's important it will be a headline" cybersecurity information seeking in older adults.
\newblock In {\em Proceedings of the 2019 CHI Conference on Human Factors in Computing Systems}, pages 1--11, 2019.

\bibitem{morrison2023recognising}
Benjamin~Alan Morrison, James Nicholson, Lynne Coventry, and Pam Briggs.
\newblock Recognising diversity in older adults' cybersecurity needs.
\newblock In {\em Proceedings of the 2023 ACM Conference on Information Technology for Social Good}, pages 437--445, 2023.

\bibitem{das2019all}
Sanchari Das, Andrew Kim, Zachary Tingle, and Christena Nippert-Eng.
\newblock All about phishing: Exploring user research through a systematic literature review.
\newblock {\em arXiv preprint arXiv:1908.05897}, 2019.

\bibitem{AAG}
Charles Griffiths.
\newblock Headline phishing statistics, 2024.
\newblock \url{https://aag-it.com/the-latest-phishing-statistics/}.

\bibitem{ellefsen2022privacy}
Jonas Ellefsen and Weiqin Chen.
\newblock Privacy and data security in everyday online services for older adults.
\newblock In {\em Proceedings of the 10th International Conference on Software Development and Technologies for Enhancing Accessibility and Fighting Info-exclusion}, pages 203--207, 2022.

\bibitem{ray2019woe}
Hirak Ray, Flynn Wolf, Ravi Kuber, and Adam~J Aviv.
\newblock " woe is me" examining older adults' perceptions of privacy.
\newblock In {\em Extended abstracts of the 2019 CHI conference on human factors in computing systems}, pages 1--6, 2019.

\bibitem{ptsb}
Permanent TSB.
\newblock Reflecting ireland, an insight into consumer behavioural change in ireland, 2022.
\newblock \url{https://www.ptsb.ie/globalassets/pdf-documents/reflecting-ireland---fraud.pdf}.

\bibitem{FraudSmart}
Banking \& Payments~Federation Ireland.
\newblock Older irish people losing almost six times more money to scammers than younger generation – fraudsmart survey, 2019.
\newblock \url{https://bpfi.ie/older-irish-people-losing-almost-six-times-money-scammers-younger-generation-fraudsmart-survey/}.

\bibitem{kemp2023consumer}
Steven Kemp and Nieves Erades~P{\'e}rez.
\newblock Consumer fraud against older adults in digital society: Examining victimization and its impact.
\newblock {\em International Journal of Environmental Research and Public Health}, 20(7):5404, 2023.

\bibitem{beh2018achieving}
Jeanie Beh, Sonja Pedell, and Bruno Mascitelli.
\newblock Achieving digital inclusion of older adults through interest-driven curriculums.
\newblock {\em The Journal of Community Informatics}, 14(1), 2018.

\bibitem{di2019psychological}
Dina Di~Giacomo, Jessica Ranieri, Meny D’Amico, Federica Guerra, and Domenico Passafiume.
\newblock Psychological barriers to digital living in older adults: computer anxiety as predictive mechanism for technophobia.
\newblock {\em Behavioral Sciences}, 9(9):96, 2019.

\bibitem{holgersson2021cybersecurity}
Jesper Holgersson, Joakim K{\"a}vrestad, and Marcus Nohlberg.
\newblock Cybersecurity and digital exclusion of seniors: What do they fear?
\newblock In {\em International Symposium on Human Aspects of Information Security and Assurance}, pages 12--21. Springer, 2021.

\bibitem{wilson2023understanding}
Gemma Wilson, Jessica~R Gates, Santosh Vijaykumar, and Deborah~J Morgan.
\newblock Understanding older adults’ use of social technology and the factors influencing use.
\newblock {\em Ageing \& Society}, 43(1):222--245, 2023.

\bibitem{feng2023understanding}
Dandi Feng, Hiba Rafih, and Cosmin Munteanu.
\newblock Understanding older adults’ safety perceptions and risk mitigation strategies when accessing online services.
\newblock In {\em International Conference on Human-Computer Interaction}, pages 467--491. Springer, 2023.

\bibitem{kottl2021but}
Hanna K{\"o}ttl, Vera Gallistl, Rebekka Rohner, and Liat Ayalon.
\newblock “but at the age of 85? forget it!”: Internalized ageism, a barrier to technology use.
\newblock {\em Journal of Aging Studies}, 59:100971, 2021.

\bibitem{barrie2021because}
Hannah Barrie, Tara La~Rose, Brian Detlor, Heidi Julien, and Alexander Serenko.
\newblock “because i’m old”: The role of ageism in older adults’ experiences of digital literacy training in public libraries.
\newblock {\em Journal of Technology in human ServiceS}, 39(4):379--404, 2021.

\bibitem{arthanat2019multi}
Sajay Arthanat, Kerryellen~G Vroman, Catherine Lysack, and Joseph Grizzetti.
\newblock Multi-stakeholder perspectives on information communication technology training for older adults: implications for teaching and learning.
\newblock {\em Disability and Rehabilitation: Assistive Technology}, 14(5):453--461, 2019.

\bibitem{gitlow2014technology}
Lynn Gitlow.
\newblock Technology use by older adults and barriers to using technology.
\newblock {\em Physical \& Occupational Therapy in Geriatrics}, 32(3):271--280, 2014.

\bibitem{bhattacharjee2020older}
Priyankar Bhattacharjee, Steven Baker, and Jenny Waycott.
\newblock Older adults and their acquisition of digital skills: A review of current research evidence.
\newblock In {\em Proceedings of the 32nd Australian conference on human-computer Interaction}, pages 437--443, 2020.

\bibitem{butt2023barriers}
Sidra~Azmat Butt, Silvia Lips, Rahul Sharma, Ingrid Pappel, and Dirk Draheim.
\newblock Barriers to digital transformation of the silver economy: Challenges to adopting digital skills by the silver generation.
\newblock {\em Proceedings of AHFE}, pages 151--163, 2023.

\bibitem{schlomann2022older}
Anna Schlomann, Christiane Even, and Torsten Hammann.
\newblock How older adults learn ict—guided and self-regulated learning in individuals with and without disabilities.
\newblock {\em Frontiers in Computer Science}, 3:803740, 2022.

\bibitem{tomczyk2023barriers}
{\L}ukasz Tomczyk, Maria~Lidia Mascia, Dorota Gierszewski, and Christopher Walker.
\newblock Barriers to digital inclusion among older people: a intergenerational reflection on the need to develop digital competences for the group with the highest level of digital exclusion.
\newblock {\em Innoeduca}, 9(1), 2023.

\bibitem{harris2022older}
Maurita~T Harris, Kenneth~A Blocker, and Wendy~A Rogers.
\newblock Older adults and smart technology: facilitators and barriers to use.
\newblock {\em Frontiers in Computer Science}, 4:835927, 2022.

\bibitem{mckee2006older}
Heidi McKee and Kristine Blair.
\newblock Older adults and community-based technological literacy programs: barriers \& benefits to learning.
\newblock {\em Community Literacy Journal}, 1(2):13--39, 2006.

\bibitem{friemel2016digital}
Thomas~N Friemel.
\newblock The digital divide has grown old: Determinants of a digital divide among seniors.
\newblock {\em New media \& society}, 18(2):313--331, 2016.

\bibitem{transport-justice}
Social~Justice Ireland.
\newblock Improving transport options in urban and rural areas, 2023.
\newblock \url{https://www.socialjustice.ie/article/improving-transport-options-urban-and-rural-areas}.

\bibitem{EU-Barometer}
European~Union Barometer.
\newblock A long term vision for eu rural areas, 2021.
\newblock \url{https://europa.eu/eurobarometer/surveys/detail/2278}.

\bibitem{flynn2024keeping}
Sandra Flynn.
\newblock Keeping up with the times in ireland: Older adults bridging the age-based digital divide together?
\newblock {\em Studies in the Education of Adults}, pages 1--19, 2024.

\bibitem{Irish-Indo-Broadband}
Adrian Weckler.
\newblock Ireland one of europe's worst for rural broadband gap, says eu report, 2021.
\newblock \url{https://www.independent.ie/business/technology/ireland-one-of-europes-worst-for-rural-broadband-gap-says-eu-report/40606621.html}.

\bibitem{hogan2006technophobia}
Mair{\'e}ad Hogan.
\newblock Technophobia amongst older adults in ireland.
\newblock {\em Irish Journal of Management}, 27(1):57, 2006.

\bibitem{pirhonen2020these}
Jari Pirhonen, Luciana Lolich, Katariina Tuominen, Outi Jolanki, and Virpi Timonen.
\newblock “these devices have not been made for older people's needs”--older adults' perceptions of digital technologies in finland and ireland.
\newblock {\em Technology in Society}, 62:101287, 2020.

\bibitem{marsick2015informal}
Victoria~J Marsick and Karen Watkins.
\newblock {\em Informal and incidental learning in the workplace (Routledge revivals)}.
\newblock Routledge, 2015.

\bibitem{topping2013trends}
Keith~J Topping.
\newblock Trends in peer learning.
\newblock {\em Developments in Educational Psychology}, pages 53--68, 2013.

\bibitem{hi-digital}
Vodafone-Alone.
\newblock About hi digital, 2023.
\newblock \url{https://hidigital.ie/about}.

\bibitem{tomczyk2022digital}
{\L}ukasz Tomczyk, Anna Mr{\'o}z, Katarzyna Potyra{\l}a, and Joanna Wn{\k{e}}k-Gozdek.
\newblock Digital inclusion from the perspective of teachers of older adults-expectations, experiences, challenges and supporting measures.
\newblock {\em Gerontology \& geriatrics education}, 43(1):132--147, 2022.

\bibitem{lobuono2020teaching}
Dara~L LoBuono, Skye~N Leedahl, and Elycia Maiocco.
\newblock Teaching technology to older adults: modalities used by student mentors and reasons for continued program participation.
\newblock {\em Journal of gerontological nursing}, 46(1):14--20, 2020.

\bibitem{schirmer2022digital}
Werner Schirmer, Nelly Geerts, Anina Vercruyssen, Ignace Glorieux, Digital~Ageing Consortium, et~al.
\newblock Digital skills training for older people: The importance of the ‘lifeworld’.
\newblock {\em Archives of gerontology and geriatrics}, 101:104695, 2022.

\bibitem{geerts2023bridging}
Nelly Geerts, Werner Schirmer, Anina Vercruyssen, and Ignace Glorieux.
\newblock Bridging the ‘instruction gap’: how ict instructors help older adults with the acquisition of digital skills.
\newblock {\em International Journal of Lifelong Education}, 42(2):195--207, 2023.

\bibitem{gates2022role}
Jessica~R Gates and Gemma Wilson-Menzfeld.
\newblock What role does geragogy play in the delivery of digital skills programs for middle and older age adults? a systematic narrative review.
\newblock {\em Journal of Applied Gerontology}, 41(8):1971--1980, 2022.

\bibitem{vercruyssen2023basic}
Anina Vercruyssen, Werner Schirmer, Nelly Geerts, and Dimitri Mortelmans.
\newblock How “basic” is basic digital literacy for older adults? insights from digital skills instructors.
\newblock In {\em Frontiers in Education}, volume~8, page 1231701. Frontiers Media SA, 2023.

\bibitem{chiu2019help}
Ching-Ju Chiu, Wan-Chen Tasi, Wan-Lin Yang, and Jong-Long Guo.
\newblock How to help older adults learn new technology? results from a multiple case research interviewing the internet technology instructors at the senior learning center.
\newblock {\em Computers \& Education}, 129:61--70, 2019.

\bibitem{cso2019womenmen}
Central Statistics~Office (CSO).
\newblock Women and men in ireland 2019 - education.
\newblock \url{https://www.cso.ie/en/releasesandpublications/ep/p-wamii/womenandmeninireland2019/education/}, 2019.
\newblock Accessed: 2024-07-31.

\bibitem{Braun2019}
Virginia Braun and Victoria Clarke.
\newblock Reflecting on reflexive thematic analysis.
\newblock {\em Qualitative Research in Sport, Exercise and Health}, 11(4):589--597, 2019.

\bibitem{braun2012thematic}
Virginia Braun and Victoria Clarke.
\newblock {\em Thematic analysis.}
\newblock American Psychological Association, 2012.

\bibitem{ahmad2022effectiveness}
Nahdatul~Akma Ahmad, Muhammad~Fairuz Abd~Rauf, Najmi~Najiha Mohd~Zaid, Azaliza Zainal, Tengku~Shahrom Tengku~Shahdan, and Fariza~Hanis Abdul~Razak.
\newblock Effectiveness of instructional strategies designed for older adults in learning digital technologies: a systematic literature review.
\newblock {\em SN computer science}, 3(2):130, 2022.

\bibitem{barros2021circumspect}
Belen Barros~Pena, Rachel~E Clarke, Lars~Erik Holmquist, and John Vines.
\newblock Circumspect users: Older adults as critical adopters and resistors of technology.
\newblock In {\em Proceedings of the 2021 CHI Conference on Human Factors in Computing Systems}, pages 1--14, 2021.

\bibitem{tilda-lonely}
Rose Anne~Kenny Mark~Ward, Richard~Layte.
\newblock Loneliness, social isolation, and their discordance among older adults, 2019.
\newblock \url{https://tilda.tcd.ie/publications/reports/pdf/Report_Loneliness.pdf}.

\bibitem{marston2019older}
Hannah~Ramsden Marston, Rebecca Genoe, Shannon Freeman, Cory Kulczycki, and Charles Musselwhite.
\newblock Older adults’ perceptions of ict: Main findings from the technology in later life (till) study.
\newblock In {\em Healthcare}, volume~7, page~86. MDPI, 2019.

\bibitem{neves2021digital}
Barbara~Barbosa Neves and Geoffrey Mead.
\newblock Digital technology and older people: Towards a sociological approach to technology adoption in later life.
\newblock {\em Sociology}, 55(5):888--905, 2021.

\bibitem{morrison2021older}
Benjamin Morrison, Lynne Coventry, and Pam Briggs.
\newblock How do older adults feel about engaging with cyber-security?
\newblock {\em Human behavior and emerging technologies}, 3(5):1033--1049, 2021.

\bibitem{rosales2022explicit}
Andrea Rosales and Daniel Blanche-T.
\newblock Explicit and implicit intergenerational digital literacy dynamics: How families contribute to overcome the digital divide of grandmothers.
\newblock {\em Journal of Intergenerational Relationships}, 20(3):328--346, 2022.

\bibitem{NBI}
National~Broadband Ireland.
\newblock Welcome to national broadband ireland, 2024.
\newblock \url{https://nbi.ie/}.

\bibitem{Irish-examiner}
Cianan Brennan.
\newblock Failure to connect: How has the national broadband plan crashed so badly?, 2021.
\newblock \url{https://www.irishexaminer.com/opinion/commentanalysis/arid-40721490.html}.

\bibitem{NESC}
National Economic \&~Social Council.
\newblock Digital inclusion in ireland: Connectivity, devices \& skills, 2021.
\newblock \url{http://files.nesc.ie/nesc_reports/en/154_Digital.pdf}.

\bibitem{keogh2021growth}
Patrick Keogh.
\newblock {\em The Growth of Digitization of Retail Banking in Ireland, and the impact FinTechs’ have on the adoption and actual usage}.
\newblock PhD thesis, Dublin Business School, 2021.

\bibitem{rajput2019impact}
Sagar Singh~Kirpalsingh Rajput.
\newblock {\em The impact of online banking attributes on customer satisfaction: A study from the Irish retail banking customer perspective}.
\newblock PhD thesis, Dublin, National College of Ireland, 2019.

\bibitem{latulipe2022unofficial}
Celine Latulipe, Ronnie Dsouza, and Murray Cumbers.
\newblock Unofficial proxies: How close others help older adults with banking.
\newblock In {\em Proceedings of the 2022 CHI Conference on Human Factors in Computing Systems}, pages 1--13, 2022.

\bibitem{ronan2022directives}
Patricia Ronan.
\newblock Directives and politeness in spice-ireland.
\newblock {\em Corpus Pragmatics}, 6(2):175--199, 2022.

\end{thebibliography}

\end{document}